\documentclass[journal]{IEEEtai}

\usepackage[colorlinks,urlcolor=blue,linkcolor=blue,citecolor=blue]{hyperref}

\usepackage{color,array}

\usepackage{graphicx}

\usepackage{comment}
\usepackage{multirow}
\usepackage{algorithmic}
\usepackage{color,array}
\usepackage{graphicx}
\usepackage{listings}
\usepackage{subfigure}
\usepackage{rotating}
\usepackage{lscape}
\usepackage{verbatim}
\usepackage{hhline}
\usepackage{textcomp,booktabs}
\usepackage{xcolor,colortbl}
\usepackage{paralist}
\usepackage{textcomp}
\usepackage{wrapfig}
\usepackage{amsthm}
\usepackage{url}
\usepackage{breakurl}
\usepackage{caption}
\usepackage{algorithm}

\begin{document}

\title{Blockchain-Empowered Cyber-Secure Federated Learning  for Trustworthy Edge Computing}

\author{Ervin Moore$^\mathbf{*}$, Ahmed Imteaj$^\mathbf{*}$, \IEEEmembership{Member, IEEE}, Md Zarif Hossain,\\ Shabnam Rezapour,  and M. Hadi Amini, \IEEEmembership{Senior Member, IEEE}
\thanks{$^\mathbf{*}$ The first two authors contributed equally to this work. }

\thanks{Corresponding Author: M. Hadi Amini, amini@cs.fiu.edu \\Ervin Moore and M. Hadi Amini are with Knight Foundation School of Computing and Information Sciences, Florida International University (FIU), Miami, FL 33199, USA. They are also with the  Sustainability, Optimization, and Learning for InterDependent networks laboratory (solid lab) and ADvanced Education and Research for Machine Learning-driven Critical Infrastructure REsilience (ADMIRE) Center at FIU. (e-mails: emoor047@fiu.edu, amini@cs.fiu.edu).}

\thanks{Ahmed Imteaj and Md Zarif Hossain are with the School of Computing, Southern Illinois University. Previously, Ahmed Imteaj was with the Knight Foundation School of Computing and Information Sciences, Florida International University, Miami, FL 33199, USA.  (e-mail: imteaj@siu.edu, mdzarif.hossain@siu.edu).}

\thanks{Shabnam Rezapour is with the Enterprise and Logistics Engineering program at FIU. (e-mail: srezapou@fiu.edu).}}

\markboth{}
{Moore \MakeLowercase{\textit{et al.}}}

\maketitle

\begin{abstract}
Federated Learning (FL) is a privacy-preserving distributed machine learning scheme, where each participant’s data remains on the participants' devices and only the local model generated utilizing the local computational power is transmitted throughout the database. However, the distributed computational nature of FL creates the necessity to develop a mechanism that can remotely trigger any network agents, track their activities, and prevent threats to the overall process posed by malicious participants. Particularly, the FL paradigm may become vulnerable due to an active attack from the network participants, called a poisonous attack. In such an attack, the malicious participant acts as a benign agent capable of affecting the global model quality by uploading a{\color{black}n obfuscated}  poisoned local model {\color{black}update} to the server. This paper presents a cross-device FL model that ensures trustworthiness, fairness, and authenticity in the underlying FL training process. We leverage trustworthiness by constructing a reputation-based trust model based on agents’ contributions toward model convergence. We ensure fairness by identifying and removing malicious agents from the training process through an outlier detection technique. Additionally, we establish authenticity by generating a token for each participating device through a distributed sensing mechanism and storing that unique token in a blockchain smart contract. Further, we insert the trust scores of all agents into a blockchain and validate their reputations using various consensus mechanisms that consider the computational task.
\end{abstract}

\begin{IEEEImpStatement}
This paper presents a novel method to enhance the cybersecurity of federated learning (FL), a modern privacy-preserving distributed machine learning technique. FL's model updates sharing mechanism exposes it to poisonous attacks from malicious participants. To mitigate these threats, our approach introduces three core strategies: trustworthiness, fairness, and authenticity. Trustworthiness involves a reputation-based system to filter agents' updates, ensuring only positive contributions affect the global model. For fairness, outlier detection is used to sideline harmful participants, thereby protecting integrity of the model. Authenticity is reinforced through a unique distributed sensing mechanism that creates device-specific tokens, verified and recorded on a blockchain. This method employs smart contracts and a proof-of-work (PoW) consensus to prevent unauthorized changes. Our approach not only addresses malicious attacks but also pioneers a more transparent and reliable FL ecosystem, advancing privacy-preserving distributed machine learning. Results demonstrate robustness against different outlier schemes that appear in networks and require dropout.
\end{IEEEImpStatement}


\section{Introduction}

The expansion of Internet of Things (IoT) devices and technologies has led to a massive increase in data availability, rising concerns about data privacy. Data leakage can occur during the storage and transmission of network communications, posing challenges for both data owners and providers. Existing works on network reliability have primarily focused on centrally calculating aggregate information about participant data while preserving participants’ privacy through traditional techniques such as k-anonymity and l-diversity. However, these methods assume that network attackers have limited background knowledge, leaving data vulnerable to background knowledge attacks \cite{desai2022background} and other adversarial activities.


Modern decentralized machine learning approaches are sought after to further reduce centralized faults. McMahan et al. \cite{mcmahan2017communication} introduced a decentralized Federated Learning (FL) approach, which included a privacy-preserving distributed machine learning scheme, where each participant’s sensitive data remains locally on the device, and only the locally generated machine learning model is shared with a central aggregator. FL includes privacy-preserving mechanisms such as differential privacy or adding noise to participant data to provide additional data protection. However, the distributed computational nature of FL necessitates the development of mechanisms that remotely calculate network agents, track their activities, and prevent network threats possibly created by adversarial network behavior. Particularly, the FL  may become vulnerable to active poisoning attacks \cite{sun2021data}, where malicious participants act as benign agents and affect the global model performance through dishonest participation.

Blockchain technology introduces various decentralized mechanisms that mitigate centralized flaws. Single points of failure vulnerabilities found in centralized technologies create risks to network reliability. Blockchain's distributed ledger disallows fabrication of false updates that deviate from the main blockchain. The multiple layers of blockchain technology can improve network reliability. Block consensus mechanisms may consider participant reputation scores, performance, and historical metadata to benefit the overall blockchain. The proposed blockchain-enabled cross-device FL architecture introduces a novel approach to improve security against malicious updates, untrustworthy participants, and anomalous device behavior.

{\color{black}
\subsection{Contributions}
{\color{black}
\begin{enumerate}
\item We introduce a blockchain-enabled federated learning framework that considers decentralized reputation scores, computing resource estimates, and FL metadata of blocks to improve the FL learning process.
\item We demonstrate the advantages of combining federated learning and blockchain technology, which leverage participants' reputation scores to mitigate adversarial attacks.
\item {\color{black}We propose a blockchain-enabled federated learning architecture  with PoW consensus over FL network data. Overall outlier detection incorporates consensus mechanisms such as proof-of-elapsed-time (PoET) to ensure a secure blockchain FL learning and data storage process.}
\end{enumerate}
}
\subsection{Related Works}
{\color{black} Related works include a discussion on cost-effective FL designs \cite{luo2021cost} and consideration of resource management \cite{rahman2021smartblock}, such as considering blockchain resource budgets on the edge \cite{fan2020hybrid}, potential in physical environments in terms of sensor data and processing distributed sensor networks \cite{sezer2023ppfchain}, architectures that examine blockchain-empowered FL that further protects data privacy of participating devices \cite{zhu2023blockchain}, \cite{10242386}, \cite{wang2023blockchain}, \cite{guo2022implementation}, decentralized peer-to-peer (P2P) approaches for multi-party machine learning using blockchain and cryptographic primitives \cite{shayan2020biscotti}, blockchain approaches that reduce the total number of nodes within its blockchain network \cite{xu2020segment}, application of privacy preserving blockchains in different aspects of the Internet of Things (IoT) \cite{zhao2020privacy}, such as industrial IoT \cite{gai2019differential},  \cite{lu2019blockchain}, related mechanisms between blockchain and Distributed Trust and Reputation Management Systems (DTRMS) \cite{bellini2020blockchain}, and consideration of various blockchain consensus mechanisms and their impact on network performance \cite{lashkari2021comprehensive}.
}

\subsection{Organization}

{\color{black}
This paper is organized into four sections: in Section \textbf{II}, we introduce the proposed framework, its components, mechanisms, operations, and overall security protocols offered when combining the blockchain and FL framework. In Section \textbf{III}, we detail the experimental setup, applying the proposed framework to the NASA turbofan dataset as a regression problem, 
 followed by an elaboration of framework defenses and  results. Section \textbf{IV} concludes the paper with a discussion regarding consideration of outliers, interpretation of noisy results, and future research directions. 
}

}

\section{Proposed Framework}
\subsection{Registration, Token Generation and Distributed Sensing Mechanism}
\begin{figure*}[htb!] 
 \setlength{\belowcaptionskip}{-5pt}
    \centering 
    \includegraphics[height=13cm]{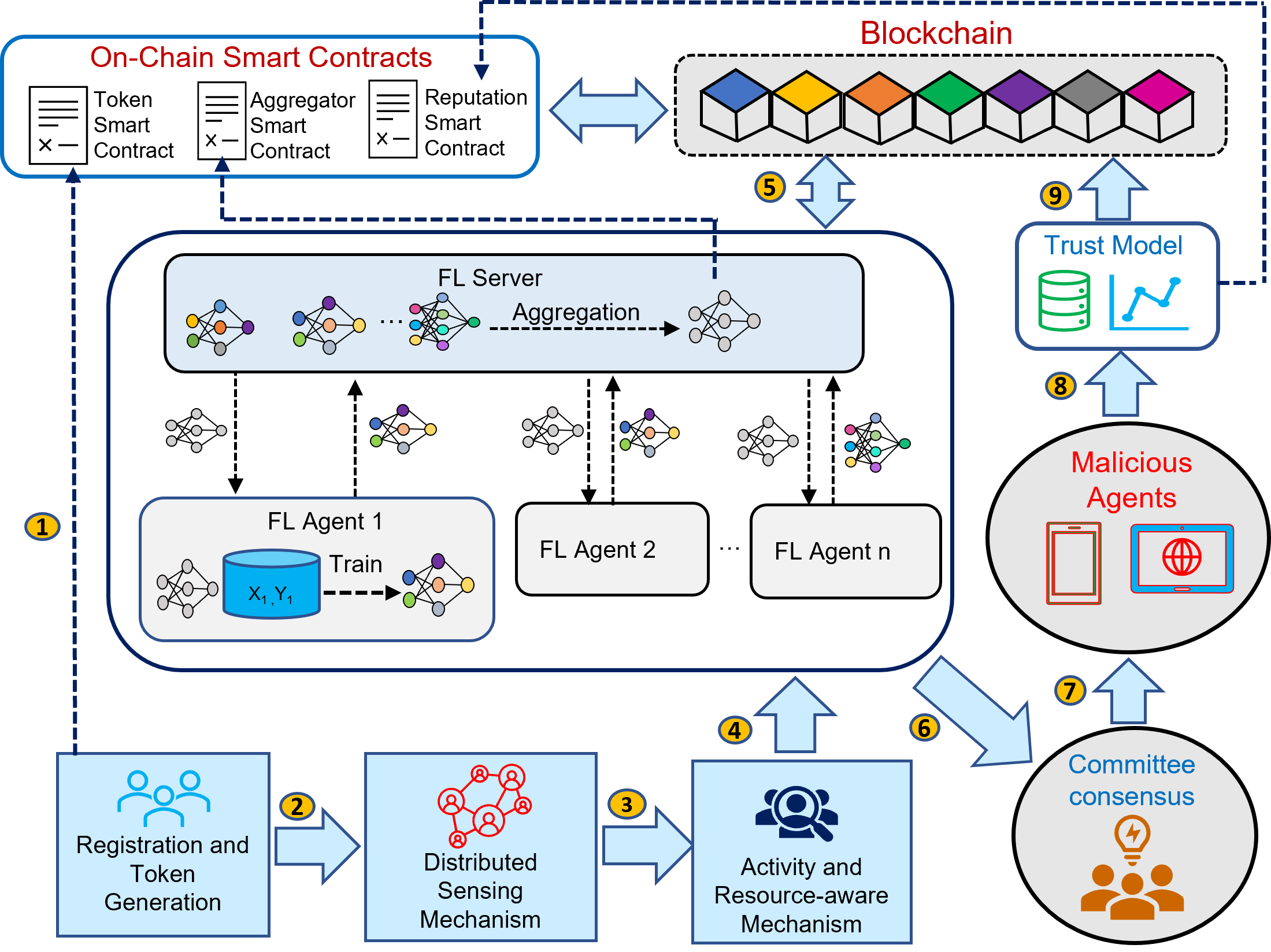} 
    \caption{Flowchart of system architecture. The framework is initialized through smart contracts, allowing participants to register, participate, and contribute towards quality FL updates. (Steps displayed are: 0) On-Chain Smart Contracts, 1) Registration and Token Generation, 2) Distributed Sensing Mechanism, 3) Activity and Resource-aware Mechanism, 5) Blockchain-FL, 6) Committee Consensus, 7) Malicious Agents, 8) Trust Model, and 9) Blockchain.)}
    \label{fig:1}
\end{figure*}

In our proposed framework (presented in Fig. \ref{fig:1}), any edge device interested in participating in the FL process needs a complete registration. Based on the registration information, we generate a unique token or identifier for each edge device. Note that, the token functions similarly to a label providing each edge device with a unique identity. These tokens are then stored inside a blockchain to ensure that the identities of the edge devices remain secure and are not tampered with by external sources. Further, we leverage a {\color{black}resource-aware} distributed sensing mechanism capable of tracking any edge device using the token and triggering a particular device by forwarding a push notification to send back required information. {\color{black} The data similarity recorded between devices with lower reputation scores can help reveal adversaries applying colluding or blockchain-based attacks such as Sybil attacks. }

\subsection{Activity and Resource-Aware Mechanism}
In the next layer of our framework, we have an activity and resource-aware mechanism capable of examining the resource status of the edge devices and monitoring their activities. {\color{black}W}e consider the {\color{black}foundational} FL approach, i.e., FedAVG \cite{mcmahan2017communication}, {\color{black}which} randomly select{\color{black}s} {\color{black}a subset of} devices to participate in the training process. However, random selection approaches do not guarantee {\color{black}the} avoidance of weak and unreliable edge devices in a highly malicious FL environment. To address this concern, we have designed a mechanism that chooses edge devices 
fulfilling the minimum resource requirements to perform a task, and evaluating the contributions of an edge device towards model convergence by analyzing its previous participation history. 
This mechanism differs from several detailed surveys on FL, where participants within networks were assumed to have unbounded resources \cite{imteaj2021surveyiot}.

{\color{black}\textbf{Resource considerations:}
The cross-platform library psutil\footnote{Psutil documentation found at https://psutil.readthedocs.io/en/latest/} is utilized to capture the resources of participants. We consider the following resources found in Table \ref{tab:Resource-considerations}. Modern operating systems such as Linux, Windows, Mac, Berkeley Software Distribution (BSD), Oracle Solaris, and IBM AIX are supported. Bandwidth is considered to verify if a device is bandwidth-constrained. Network connectivity is estimated using parameters such as packet loss to estimate how congested a user network is. Python support accommodates Python 2.7, commonly found in the majority of Linux distributions, and sometimes pre-installed. Python 3.4+ is popular in modern distributions, while PyPy is often used for optimization techniques due to its just-in-time compiling. We consider two types of devices for battery life: 1) Desktop is assumed to be plugged into a power source, 2) Mobile devices that can be moved between locations seamlessly. While a mobile device is in transport, we recognize the device is not plugged in and identify its remaining battery \%. Otherwise, we examine if the device is plugged in and if its battery \% is reasonable. Disk usage is considered due to the size of the blockchain and FL local data being stored on the edge device. CPU resources are considered for processing. Physical CPUs are measured to identify the number of CPU cores on the machine. Logical CPUs are measured to estimate the ability of these cores to handle multiple processes simultaneously given the physical CPUs. CPU utilization allows a general idea of the current workload being handled by the machine. Memory availability looks at virtual memory availability, as well as swap memory availability when available RAM is lacking.

\begin{table*}[]
\caption{Hardware considerations in resource-aware approach}
\centering
{
\begin{tabular}{|l|l|l|l|l|l|l|}
\hline
\begin{tabular}[c]{@{}l@{}}Operating system\end{tabular} & Bandwidth        & \begin{tabular}[c]{@{}l@{}}Python \\ support\end{tabular} & \begin{tabular}[c]{@{}l@{}}Battery life\end{tabular}                                 & Disk usage                & \begin{tabular}[c]{@{}l@{}}CPU \\ resources\end{tabular}                       & \begin{tabular}[c]{@{}l@{}}Memory \\ availability\end{tabular}                     \\ \hline
Linux                                                       & Bytes sent       & \multirow{2}{*}{2.7}                                      & \multirow{2}{*}{Desktop}                                                                & \multirow{2}{*}{Total GB} & \multirow{2}{*}{\begin{tabular}[c]{@{}l@{}}Logical\\ CPUs\end{tabular}}        & \multirow{3}{*}{\begin{tabular}[c]{@{}l@{}}Virtual memory\end{tabular}}   \\ \cline{1-2}
Windows                                                     & Bytes received   &                                                           &                                                                                         &                           &                                                                                &                                                                                    \\ \cline{1-6}
MacOS                                                       & Packets sent     & \multirow{2}{*}{3.4+}                                     & \multirow{2}{*}{\begin{tabular}[c]{@{}l@{}}Non-desktop \\ plugged-in\end{tabular}}      & \multirow{2}{*}{Used GB}  & \multirow{2}{*}{\begin{tabular}[c]{@{}l@{}}Physical\\ CPUs\end{tabular}}       &                                                                                    \\ \cline{1-2} \cline{7-7} 
BSD                                    & Packets received &                                                           &                                                                                         &                           &                                                                                & \multirow{3}{*}{\begin{tabular}[c]{@{}l@{}}Swap memory\end{tabular}} \\ \cline{1-6}
Sun Solaris                                                 & I/O Packet error & \multirow{2}{*}{PyPy}                                     & \multirow{2}{*}{\begin{tabular}[c]{@{}l@{}}Not plugged-in,\\ remaining \%\end{tabular}} & \multirow{2}{*}{Free GB}  & \multirow{2}{*}{\begin{tabular}[c]{@{}l@{}}\% CPU \\ utilization\end{tabular}} &                                                                                    \\ \cline{1-2}
AIX                                                         & I/O Packet drop  &                                                           &                                                                                         &                           &                                                                                &                                                                                    \\ \hline
\end{tabular}}

\label{tab:Resource-considerations} 
\end{table*}}

\subsection{On-Chain Federated Learning Components}
The first on-chain component is the token smart contract. Any interested network device can register and receive a unique token. Each time a new edge device joins the training network, its token is included in the chain. We assume that each edge device voluntarily participates in the training process considering the privacy-preserving decentralized nature of FL. They publish their resource status (e.g., processing power, battery life, memory availability, bandwidth, data volume) and QoS 
requirements (e.g., training budget, training time window, required number of edge devices for training, target accuracy).

The second component is the on-chain aggregator smart contract. Each selected edge device performs local model training and shares its updated local model weights. The aggregator smart contract carries out the aggregation of the collected model weights and appends the updated global model to the blockchain, which is then distributed to chosen participants in the next smart contract.  

The third on-chain FL component is the reputation smart contract. The main purpose of the reputation smart contract is to keep the trust scores of edge devices within the blockchain to improve the fairness of the participating edge devices' selection process. During the FL process, a bi-directional trust is established between FL data distributors and participating FL edge devices. As such, the FL server is responsible for monitoring the activities of the edge devices to discover potential contributors to model convergence. Increasing the reputation score of participants providing quality updates can motivate subsequent training.

\begin{figure*}[h!] 
\setlength{\belowcaptionskip}{-15pt}
    \centering 
    \includegraphics[width=0.94\linewidth]{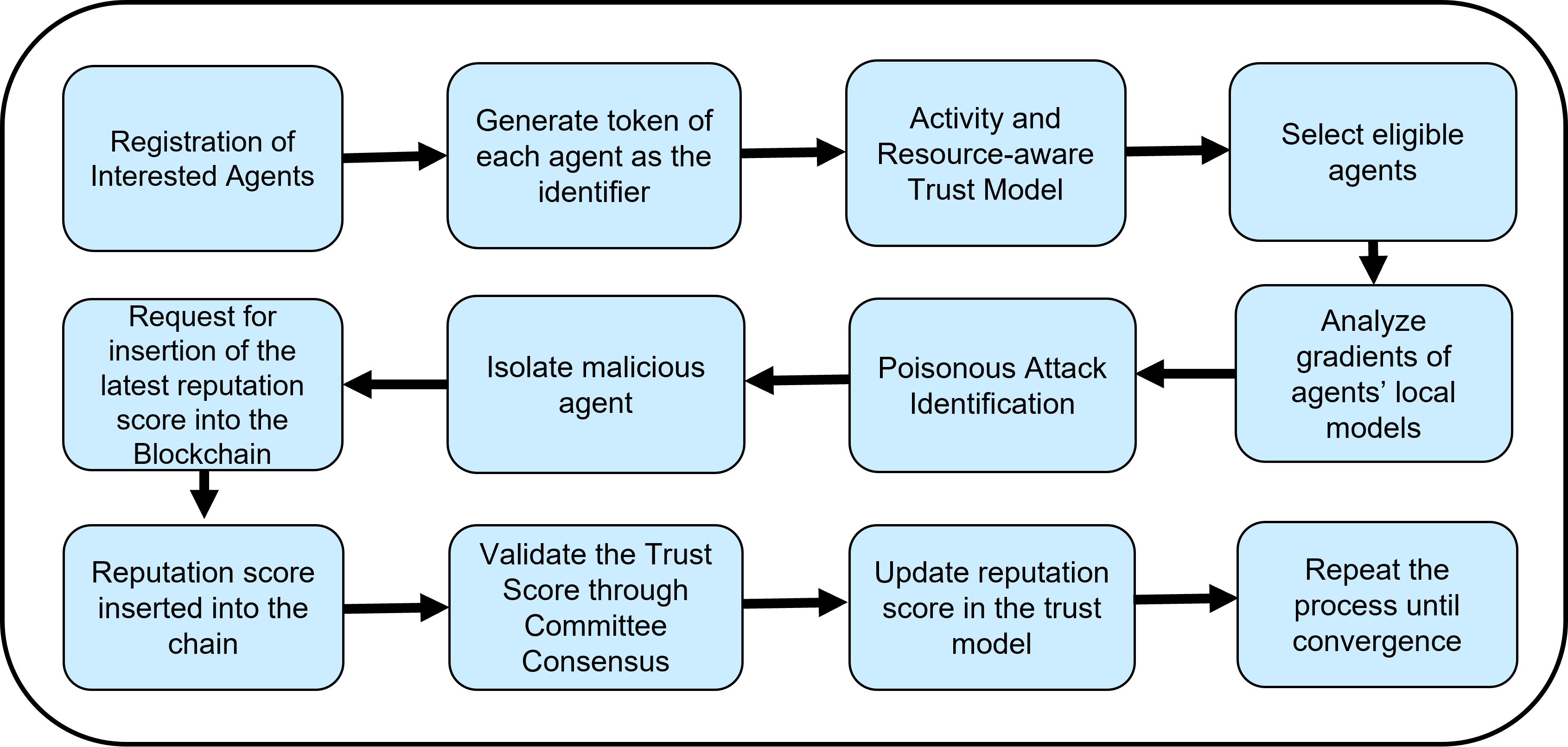} 
    \caption{ Workflow of our proposed framework starts with the registration of interested agents and continually loops through different blockchain network defense mechanisms until FL convergence. Prior works may not have considered FL participant resources, eligibility, and network reputation. Network defense mechanisms include the assessment of local FL model updates to identify malicious agents or outliers performing data/model poisoning attacks.}
    \label{fig:2}
\end{figure*}

\subsection{Off-chain Operations}\label{offchain}
Off-chain operations include the calculations of resources, reputation scores, and consensus. The update of edge devices' reputation scores through  accumulation of their previous scores is executed off-chain. This process involves analyzing the activities of the participating edge devices and considering various events and their corresponding reputation score. For instance, we consider several crucial events during an FL process, such as interested edge device, response delay, successful model update, inability to participate due to limited resources, and divergent model updates. To identify divergent or improper model parameter infusion,  we applied statistical methods for outlier detection. If an honest edge device responds with a fair model update in a timely fashion, we increase its reputation score and decrease the reputation score of the malicious and straggler edge devices. {\color{black}Computing overhead is the local training, aggregation, poisonous attack identification, and poisonous attack countermeasures such as committee consensus. The storage overhead includes on-device storage such as local blocks or local model storage. }

\begin{figure}[h!]
\setlength{\belowcaptionskip}{-15pt}
\resizebox{9cm}{!}{
    \includegraphics{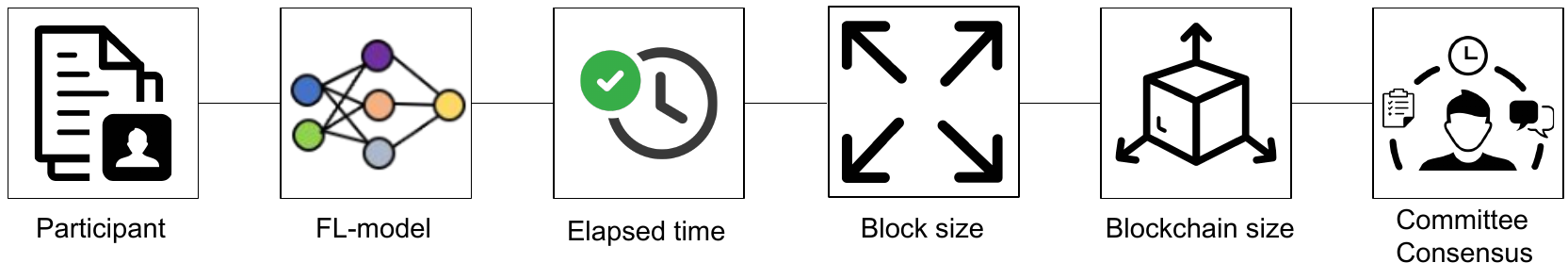}
    }

    \caption{Representation of network data stored within each block of the blockchain. Each of the six subsections is combined within a validated block.}
    \label{fig:3}
\end{figure}

\subsection{On-device Training and Generate Local Models}\label{ondevice}
On-device FL training involves a collaborative process where participating devices receive the prior blocks global model, and attempt to improve this models performance given available local data. Each FL process starts with a copy of the prior global model that was distributed to edge devices. Each device independently performs local training on its local data, optimizing the model parameters while preserving privacy. After completing local device training, participants send their local models back to the FL data distributors for aggregation. FL data administrators can combine the model updates from multiple devices using secure (SecAgg \cite{bonawitz2017practical}) or non-secure \cite{mcmahan2017communication} aggregation methods. Following the aggregation, the server generates an updated global model which gets appended to the blockchain. Once the global model is approved by the consensus mechanism and blockchain, the global model is then redistributed to participating devices. This repeated process allows a distribution of devices to continually train on their local data securely. Finally, at aggregation, quality model updates are transmitted to improve network performance.

\subsection{Identification of Poisonous Attack and Isolate Malicious Agent}
{\color{black} 
Poisonous attacks \cite{tolpegin2020data,sun2021data} refer to the intentional injection of malicious data or perturbation of training data and labels to compromise the integrity or performance of ML models. Utilizing perturbed training data or labels during the model's learning phase leads to inaccurate predictions from the trained model. Various techniques can be employed to detect and mitigate such attacks. One technique involves isolating malicious agents by analyzing their
 local updates \cite{hossain2023flid}. If a participant's local model is trained on perturbed data, that local model should appear abnormal and significantly diverge from updates produced by participants with unperturbed data. To identify such anomalous updates, the Euclidean distance between the client's weight parameters is calculated to quantify the dissimilarity among updates. Subsequently, the K-means clustering algorithm further groups the updates into distinct clusters. While local model Updates are being compared, perturbed client updates typically congregate in one cluster, while updates from unperturbed clients form a separate cluster due to their substantial deviations. The less-dense and smaller cluster is defined as the ``outlier cluster" by the standard deviation metric and is excluded in the subsequent FL aggregation phase. By discarding this cluster, we ensure that potentially compromised or perturbed updates do not influence the aggregation process and overall model performance.

 {\color{black} 
 This approach involves isolating anomalous local updates through the calculation of Euclidean distance and leveraging K-means clustering to segregate perturbed updates into a distinct cluster, which is subsequently excluded during the aggregation phase. By discarding this “outlier cluster,” we ensure that potentially anomalous updates do not influence the aggregated FL model, effectively mitigating the impact of data poisoning. The reputation-based trust model ensures efficient participant selection, reducing the computational overhead in environments with a high number of edge devices. Additionally, by leveraging local blockchains for storage and computation, we significantly reduce the data transmission and storage requirements, making the system adaptable to resource-constrained participants in large-scale settings.}

{\color{black} In addition to identifying and isolating poisonous updates, the proposed framework integrates gradient obfuscation to address membership inference attacks. Such attacks exploit gradients sent during training to infer whether a specific data point belongs to a participant's training dataset. To mitigate this, the gradients are perturbed using Gaussian noise, as defined by:

$$
\tilde{g}_i=g_i+\mathcal{N}\left(0, \sigma^2\right)
$$

where $g_i$ is the original gradient for client $i$, and $\mathcal{N}\left(0, \sigma^2\right)$ represents Gaussian noise with mean 0 and variance $\sigma^2$. This technique ensures differential privacy, as the Gaussian mechanism satisfies $(\epsilon, \delta)$-differential privacy if:

$$
\epsilon \leq \frac{\Delta}{\sigma} \sqrt{2 \ln \frac{1.25}{\delta}}
$$

Here, $\Delta=\max _{g_i, g_j}\left\|g_i-g_j\right\|_2$ is the sensitivity of the gradients. By setting $\sigma$ appropriately, $\epsilon$ can be made sufficiently small to achieve privacy while maintaining acceptable model accuracy. This gradient obfuscation technique ensures that individual gradients do not leak sensitive information about participants' training data, thereby defending against membership inference attacks and enhancing the security of the FL process. }
}

\begin{figure}[h!]
\setlength{\belowcaptionskip}{-15pt}
\resizebox{8cm}{!}{
    \includegraphics{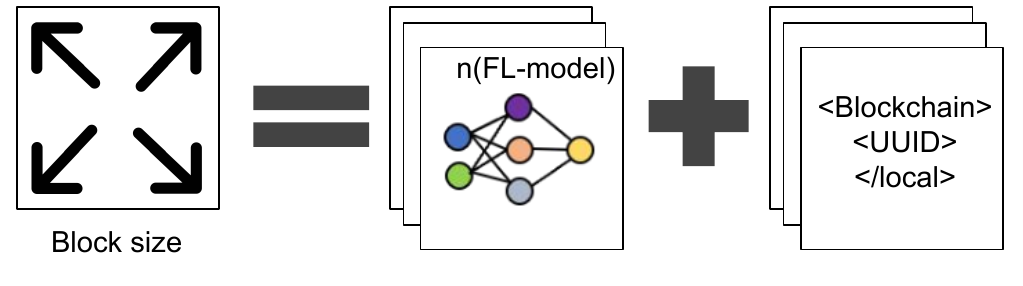}}
 
    \caption{Zoomed in the representation of transmitted block size.}
    \label{fig:4}
\end{figure}

\subsection{Leveraging Committee Consensus to Validate Model Update}
The blockchain and its block cryptographic structure guarantee the immutability of stored block information. Thus, before appending any block to the blockchain, it needs to be properly validated. The existing approaches such as, competition-based consensus methods append blocks to the existing chain and verify consensus fruition later. In contrast, communication-based consensus mechanisms reach an agreement before including any new block to the chain. However, such consensus mechanisms encounter communication and computational costs. Considering these costs, we leverage a secure and efficient Committee Consensus Mechanism (CCM) that can validate the agents' local gradients and include the correct ones in the chain. Under the CCM setting, a committee is formed with a few trustworthy agents responsible for setting smart contract thresholds for participating agents' local gradient validation and block generation. Apart from the committee nodes, the rest of the participating edge devices perform on-device training and share the{\color{black}ir local model updates for consensus consideration}. The committee examines the model quality and updates 
 {\color{black}s} the trust scores accordingly. The consensus committee is responsible for appending only the qualified local model updated onto the blockchain. In the subsequent training round, the committee is updated based on the edge devices' trust scores from the previous training round, disregarding the previous consensus committee. The consensus committee performs validations of the local gradients by considering their local data samples as a validation set and assigns scores based on the validation accuracy. This approach only requires the committee to run the model without any additional operations. After accumulating the scores of chosen committee members, the median committee score is adopted as the score of the agents' local model update. {\color{black}By applying a committee consensus mechanism similar to \cite{li2020blockchain}, we can recognize malicious devices that deviate from consensus standards. {\color{black} Anomalies are calculated by consensus using the committee consensus scores and mechanisms.}}

\subsection{Insert Valid Model update and Reputation Score into Blockchain}

{\color{black}
A valid model update includes quality data that has been screened for outlier updates. These updates improve the predictive capabilities of the global FL model through sustainable learning and robustness. Furthermore, valid model updates contribute to a more efficient FL model, potentially reducing storage and computational requirements. By validating each participant's model update we mitigate the risk of dishonest participants from compromising network resources through adversarial attacks such as poisonous attacks.

Each FL participant is assigned their own local blockchain. Each local blockchain contains the user's local default reputation score, resource requirements for participation, and smart metadata that may carry over from previous consensus. After each successful valid model update or FL aggregation, the reputation score is adjusted with the created block. Each created local blockchain block follows the structure as: \textbf{[Participant identifier] | [FL-model-x] | [Total time taken] | [block size] [blockchain-FL size] [CC-score]}. Architecture parameters and metadata are considered within the  \textbf{[CC-score]} mechanism or off-chain operations, referenced in Section \ref{offchain} The \textbf{[blockchain-FL size]} includes all network technologies for operation (fl-technologies, local requirements, and blockchain metadata). While the \textbf{[block size]} only includes previous FL models and pointers to the blockchain architecture. 
}

\begin{algorithm}
\caption{Block Validation}
\label{alg:block_validation}
\begin{algorithmic}[1]
\REQUIRE $x$ participants, data requirements, standard deviation ($\sigma$) for outliers, genesis block
\STATE \textbf{function} successfulAggregation
    \IF{$x$ users eligible for participation}
        \STATE a) Add participant to FL process
        \STATE b) Download FL parameters from prior block
        \STATE c) Process FL data locally
        \STATE d) Send FL results to aggregator
    \ENDIF
    \IF{aggregation received}
        \STATE a) Compare results to previous aggregation
        \STATE b) Remove outliers, lower participation score
        \STATE c) Save successful aggregation to block
    \ENDIF
    \IF{block received}
        \STATE a) Apply consensus to block
        \STATE b) Append block to blockchain
    \ENDIF
\STATE \textbf{end function} 
\end{algorithmic}
\end{algorithm}
\subsection{Problem Definition}
{\color{black}Centralized databases may undergo multiple challenges such as the handling of an abundance of data, data storage considerations, and the potential for re-purposing data towards optimal computing problems. Distributed computing challenges include 1) data heterogeneity, arising from variations in data samples, sources, and types; 2) stemming from the diverse hardware, networking, and resource capacities among participants; 3) Concerns regarding dishonest participants potentially undermining network performance. {\color{black} Different consensus mechanisms can affect the security and threat of a distributed computing environment  environment. Consensus mechanisms can assist in mitigating specific threats, such as Sybil attacks, collusion, and tampering that may occur during the aggregation process.}
Many distributed computing architectures encounter similar challenges as those in online computing environments. Decentralized databases like blockchain offer a solution to mitigate vulnerabilities inherent in centralized networks, such as the risk of a single point of failure. Moreover, the proposed blockchain system, which dynamically considers participant resources in real-time, can assess participant delays resulting from resource constraints. Prior works have overlooked participant resources leading to scenarios where participants with limited storage space might prematurely halt training, thereby causing the FL update process to stall and ultimately be discarded. {\color{black} Participants that cause persistent delays in aggregation receive lower consensus scores, low scores can eventually lead to removal of participation. A participant undergoing a side channel attack that lowers the servers performance is grouped with  poisonous attacks within our defenses. We categorize potential threats into three primary types: 

\begin{enumerate}

\item External Attacks: these include adversaries attempting to intercept or alter data-in-transit, which attempt to externally disrupt the blockchain’s immutable ledger and encryption techniques. 
\item Internal Attacks: These cover malicious participants introducing poisoned updates or colluding to disrupt the global model. The reputation-based scoring and outlier detection mechanisms in our framework are designed to isolate such threats and deter dishonest behavior. 
\item Physical or Side-Channel Attacks: These attacks include information leakage caused by physical hardware or system information leakage. Users purposely lowering participation scores result in consensus score penalties and eventual exclusion from participation.
\end{enumerate}
}

}

\section{Experimental Evaluation}
\subsection{Experimental Setup}
{\color{black}The experimental setup consists of a FL environment processing NASA turbofan dataset. The dataset is processed as a regression problem. We look at multiple sensor measurements gathered from one operational setting. Around 20,000x10 training data samples are utilized in one operational setting. The NASA turbofan dataset was chosen due to the complexity of sensor networks (environmental factors). In this scenario, an outlier represents a malfunctioning device update that requires dropout and device diagnosis. 

{\color{black}

}
\subsection{Results and Discussion}

System robustness was a focus of parameter tuning. Experiment training included a distribution of training data, where each distribution contained randomly generated amounts of outliers to prevent model overfitting. Random data was then amplified substantially to appear as malfunctioning device readings during training. Each outlier detected during local training is evaluated, leading to potential dropout, regularization and consensus. Increased noise levels lead to increased outlier evaluations within the network.

\begin{figure}[htb!] 
\setlength{\belowcaptionskip}{-12pt}
    \centering 
    \includegraphics[height=5.2cm]{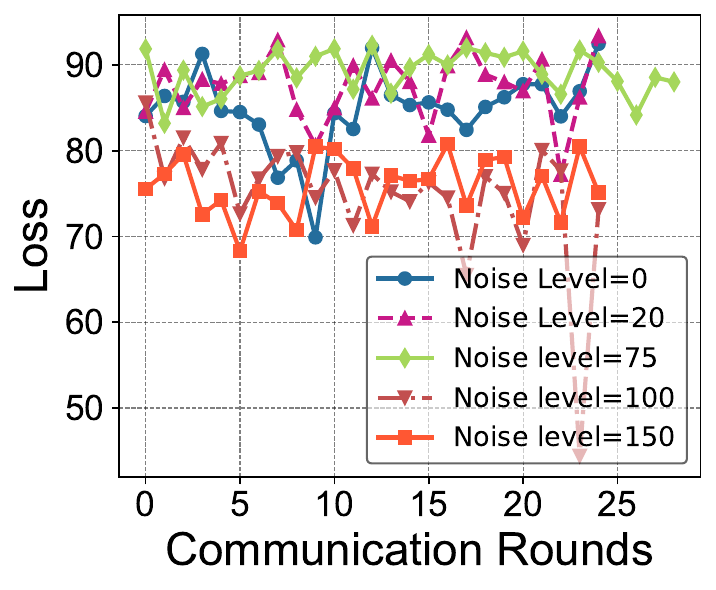} 
    \caption{Loss minimization results with the various noise levels: 0, 20, 75, 100, and 150. Noise did not ultimately lead to longer training times, as the model would actively remove detected noise and outliers to improve overall performance.} 
    \label{fig:5}
\end{figure}

\begin{figure}[htb!] 
\setlength{\belowcaptionskip}{-12pt}
    \centering 
    \includegraphics[height=5cm]{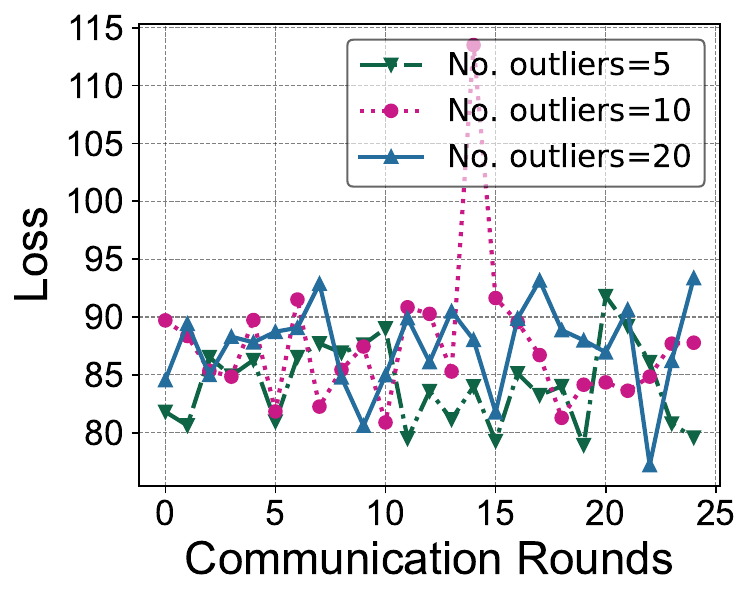} 
    \caption{Loss minimization results with different numbers of outliers. The amount of outliers in training that resulted in dropout and regularization boosted performance. In some cases the reintroduction of outliers leads to a loss climax.} 
    \label{fig:6}
\end{figure}

{\color{black} FL results in NASA turbofan regression data show learning during the first half of training and convergence in the second half of training. As demonstrated in Fig. \ref{fig:1}, the FL model eventually converges. Potential solutions for this convergence can be represented as a cryptographic problem, where optimal smart contract parameters may offer optimization. For example, reverting to an earlier saved model that demonstrated exemplary performance can be considered. {\color{black} To reduce the total FL communication costs, communications can be quantized with the federated trained ternary quantization (FTTQ) algorithm found in \cite{xu2020ternary}, which optimizes the quantized networks on the clients through a self-learning quantization factor \cite{xu2020ternary}.}

Fig. \ref{fig:6} suggests that the number of outliers is not as significant as the weights of the outliers. Randomly generated outlier weights were found to decrease overall training performance at random. The similarity of the outlier data corresponded to steep learning curves. {\color{black} Considering a data selection policy such as \cite{gong2023store} can help relieve limited device storage requirements. }

}

\begin{figure}[htb!] 
\setlength{\belowcaptionskip}{-12pt}
    \centering 
    \includegraphics[height=5cm]{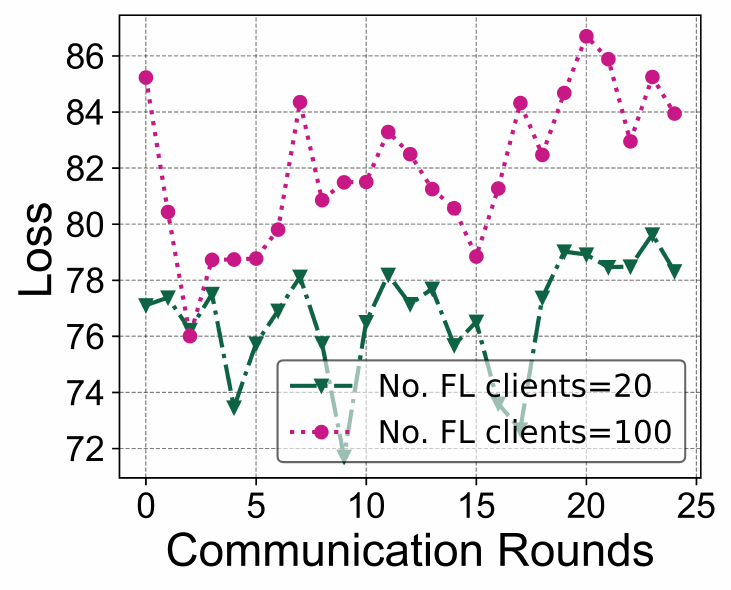} 
    \caption{Loss minimization results with various amounts of FL clients. The number of clients and total time required for aggregation increased collectively, where 100 clients took twice as long as 10.} 
    \label{fig:7}
\end{figure}

When building the FL model and tuning successfully aggregated FL models towards blockchain, we found less strict parameters had higher generalization and were less prone to overfitting. The assumption was FL outliers were stored in a “strict” format, then the strict model focused on outliers over data findings from previous models. {\color{black} To preserve data confidentiality and integrity, it is necessary to store encrypted data in the cloud storage servers.  According to Huang et al \cite{huang2014survey}, data deduplication is a specialized data compression technique that identifies common data chunks within and across different files, and stores them only once to improve storage utilization \cite{huang2014survey}. Data compression was considered within our results, and displayed in Figure \ref{fig:8}.}
\begin{figure}[htb!] 
\setlength{\belowcaptionskip}{-12pt}
    \centering 
    \includegraphics[height=5cm]{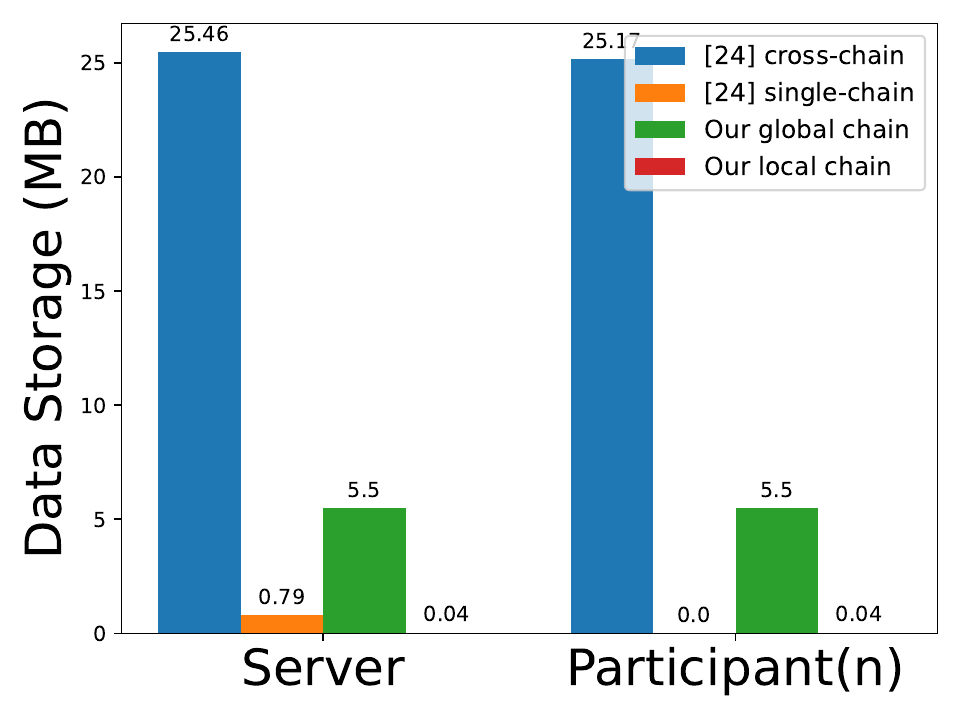} 
    \caption{Data storage comparison between our framework and a similar blockchain-empowered FL architecture applied to healthcare metaverses \cite{kang2023blockchain}.} 
    \label{fig:8}
\end{figure}

{\color{black}
{\color{black}Fig. \ref{fig:8} presents a bar chart comparative analysis of data storage requirements for both a central server and participants in an FL system, under various blockchain configurations.  For the server, it shows a significant 25.46 MB storage requirement when employing a cross-chain method as per a referenced study \cite{kang2023blockchain}, and 5.5 MB for a single-chain method from the same study. In contrast, the proposed framework's local chain requires a notably lower storage of 0.79 MB, with no specific data provided for their global chain. On the participant side, the chart illustrates a stark difference: both the cross-chain and single-chain methods from study \cite{kang2023blockchain} demand no storage from participants, while our framework requires a mere 0.04 MB for the local chain and none for the global chain. This comparison highlights the efficiency of our proposed framework, particularly the local chain approach, in reducing storage demands on the server and participants, which can be especially advantageous in sectors like smart city infrastructure \cite{bawany2015smart}, where efficient data storage and management are critical for handling the vast amounts of data generated by urban sensors and IoT devices.

A closer look at Fig. \ref{fig:8} shows participant block sizes equal to .03MB + .01MB(n), mentioned in Fig. \ref{fig:4}. While, the blockchain FL size was equal to 5.5MB + .01MB(n), containing parameters from Fig. \ref{fig:3}. Parameter ($n$) is the current iteration of the continual process, while the .01MB(n) is the additional block metadata gained during successful block aggregation. In comparison, the modern blockchain peer-to-peer network, Bitcoin Core, includes a protocol that limits blocks to 1MB in size \cite{8215367}. Results suggest current block size and storage referenced in Fig. \ref{fig:3} is minimal. Further, the FL-model[n] of each block contains the prior block model. Each model is saved into a Pytorch .pt file. Models are loaded within a parameter dictionary, where parameters are not strict with returned keys differing from input keys. The distributed learning over FL-model[n] is similar to FLchain \cite{majeed2019flchain}, a Blockchain FL architecture that trains distributed FL models on different channels for multi-access edge computing.

A discussion can be made about the application of this blockchain FL framework and its opportunity to} connect with similar frameworks. This framework's interoperability starts with the simplicity of generated blocks. The blockchain is converted into .csv format, where each block's metadata is in .txt form. If the FL or blockchain architecture is disrupted, the .txt metadata can restore the blockchain to its previous state. Thus adjacent blockchains could provide FL participants, resources, and metadata. External blockchain technologies may also alleviate some of the blockchain FL network strains, such as potential participants, resources, and networking. {\color{black} Blockchains including archival data, could refer to network storage systems, such as \cite{quinlan2002venti} to assist in data storage when the blockchain is resource-exhausted.}}
}
{\color{black}Additionally, launching attacks on tabular data posed a significant challenge due to the absence of direct pixel-level manipulation as in image data. Traditional adversarial attack methods reliant on pixel perturbations were not directly applicable, necessitating the exploration of novel techniques tailored for tabular datasets. {\color{black} During our experiments, our dataset dimensions were static and experimental data was randomly distributed, future works could  consider a flexible dataset learning from data streams \cite{marfoq2023federated}, which can increase data considerations. Incorporating advanced adversarial detection techniques, such as anomaly detection powered by machine learning, and exploring hybrid consensus mechanisms to balance security with computational efficiency are all future directions.}
}

\section{Conclusion and Future Directions}
{\color{black}
In conclusion, we proposed a robust blockchain-enabled FL architecture that considers participant resources for preferable network participation. During training, participants abnormal data is highlighted as an outlier, calculated as an outlier, then undergoes defense mechanisms (regularization or dropout) before aggregation. We then applied the proposed blockchain FL framework to a regression problem with consideration of outliers, adversaries, and noise. Participant resources were gathered to estimate potential and effectiveness of FL participation. The blockchain FL framework supports local device registration, local block/blockchain generation, local metadata, and consensus mechanisms to ensure network reliability. We demonstrate the capabilities of blockchain-enabled FL by examining each layer's mechanisms, requirements, and defenses.
Results demonstrate robust performance under various conditions of noise, clients, and outliers. Gathering resources of eligible FL participants assured preferable participation standards were upheld. Each block that was generated in the distributed blockchain was unique and authenticated, reducing misuse such as tampering. Block updates that were transmitted throughout the blockchain-empowered FL network underwent various consensus mechanisms found within PoW and PoET.

{\color{black}
Future research directions include networking costs, best consensus approaches, scalability and interoperability. Further consideration into online networking costs, such as the tradeoff between throughput and block size can be conducted. Frameworks, such as \cite{8215367}, \cite{wang2022platform} examined the communication costs of blockchains.} The number of layers within a blockchain can increase communication costs. The construction of optimal resource allocations can reduce network costs. Adaptive blockchain sharing approaches introduced in \cite{lin2023drl} can reduce the costs associated with uneven blockchain shards that may emerge. Latency between the blockchain main-chain and side-chains can be concerning. The costs of consensus can create miscommunication if blocks are generated at the same time. Different consensus mechanisms can be considered to reduce costs (PFBT, PoA, PoI, PoC and PoB) \cite{imteaj2021foundations}.  Blockchain communication limits are seen in existing blockchain technologies, where transactions per second (tps) have a maximum in terms of scalability. Chauhan et al. \cite{chauhan2018blockchain} states how ``a bitcoin scalability issue arose with the old POW consensus method and bitcoin peak limit of processing only 7 transactions per second \cite{chauhan2018blockchain}''. Multidimensional blockchain approaches such as \cite{zangoti2022multidimensional}, may also alleviate blockchain scalability issues. Lastly, blockchain interoperability and its impact on the future of blockchain technologies is of interest. A blockchain lacking resources in real-time could benefit from a neighboring blockchain transferring network resources using blockchain interoperability \cite{10242386}.

}

\section{{acknowledgment}}
Ervin Moore and M. Hadi Amini's work is partly supported by the U.S. Department of Homeland Security under  Grant Award Number 23STSLA00016-01-00.
The views and conclusions contained in this document are those of the authors and should not be interpreted as necessarily representing the official policies, either expressed or implied, of the U.S. Department of Homeland Security.

This work is also based upon the work partly supported by the National Center for Transportation Cybersecurity and Resiliency (TraCR) (a U.S. Department of Transportation National University Transportation Center) headquartered at Clemson University, Clemson, SC, USA. Any opinions, findings, conclusions, and recommendations expressed in this material are those of the author(s) and do not necessarily reflect the views of TraCR, and the U.S. Government assumes no liability for the contents or use thereof.

\bibliographystyle{unsrt}
\bibliography{refers}

\end{document}